\documentclass[12pt,preprint]{aastex}

\slugcomment{accepted for publication in the Astrophysical Journal}

\shorttitle{LBGs at z$>$5}
\shortauthors{Lehnert and Bremer}

\begin{document}

\title{Luminous Lyman Break Galaxies at z$>$5 and the Source of
Reionization\altaffilmark{1}}

\author{Matthew D. Lehnert\altaffilmark{2} and Malcolm Bremer\altaffilmark{3}}

\altaffiltext{1}{Based on observations collected at the Very Large
Telescope part of the European Southern Observatory, Cerro Paranal, Chile}
\altaffiltext{2}{Max-Planck-Institut f\"ur extraterrestrische Physik,
Giessenbachstra\ss e, 85748 Garching bei M\"unchen, Germany}
\altaffiltext{3}{Department of Physics, Bristol University, H. H. Wills
Physics Laboratory, Tyndall Avenue, Bristol, BS8 1TL, U.K.}

\begin{abstract}

We have discovered six galaxies with spectroscopically confirmed
redshifts of $4.8<z<5.8$ in a single 44 square arcminute field imaged
deeply in $R,I$ and $z-$bands. All the spectra show an emission-line
in the region around 7000-8400\AA \ with a spectroscopically-detected
faint continuum break across the line. These six were drawn from 13
sources with $I_{AB}<26.2$ and $R_{AB}-I_{AB}>1.5$ in the field,
this photometric cut designed to select galaxies at $z>4.8$. The
line fluxes range between 0.2 to 2.5$\times$10$^{-17}$ ergs cm$^{-2}$
s$^{-1}$ indicating luminosities of around 10$^{42-43}$ ergs s$^{-1}$
for Ly$\alpha$ and their high emission line equivalent widths suggest
very young ages ($\la$10$^8$ yrs).  A further line-emitting object with
no detectable continuum was serendipitously detected by spectroscopy. If
this line is Ly$\alpha$ then it is from a source at $z=6.6$, making this
the most distant galaxy known. However, the redshift cannot be considered
secure as it is based on a single line. No broad emission line objects
(quasars) were detected. The 13 sources at $I_{AB}<26.2$ are less than
that expected if the luminosity function of dropout galaxies remained
unchanged between $z=3$ and $z=6$, although the deficit is not highly
significant given possible cosmic variance.  The UV luminosity density
from galaxies brighter than our flux limit is considerably less than
that necessary to keep the volume probed by our field at $<z>\sim 5.3$
ionized. These galaxies are observed within several hundred Myr of the
end of the epoch of reionization ($z=6-7$), with little time for the
luminosity function to evolve. This, and the lack of detected quasars,
imply that the bulk of the UV flux that reionized the universe came from
faint galaxies with $M_{AB}(1700\AA)>-21$.

\end{abstract}


\section{Introduction}

While progress has been made in our understanding of how galaxies formed and
evolved by studying the dynamics, morphologies, and stellar populations of
low redshift galaxies, a complete physical picture can only be obtained by
actually witnessing the important physical processes in situ.  With this
in mind there has been an explosion in the number of galaxies and QSOs
discovered at the highest redshifts \citep[z$>$5; e.g.,][]{Hu02, Ellis01,
SS99, Dey98, Fan01, Rhoads02}.  These galaxies have been found using a
variety of techniques ranging from narrow-band imaging, to accidental
alignments of spectroscopic slits on high redshift objects, to using the
drop-out technique (Lyman Break galaxies -- LBGs) pioneered by Steidel
and collaborators.  Such a diversity of approach is necessary in order
to make a full census of the high redshift population.  Without such a
census, it is impossible to elucidate the physical processes that drive
galaxy evolution at the highest redshifts.  Moreover, to find the objects
that may have reionized the universe probably means that we will have
to obtain a full luminosity function of the most plausible objects such
as QSOs, massive stars in galactic fragments, or perhaps even something
more exotic \citep[e.g.,][]{LBrev01}.

In the present paper, we describe a program to find faint
continuum-emitting objects at z$\approx$5.5.  Our hope was to detect
objects at an epoch close enough to that of reionization so that we could
determine whether UV luminous galaxies or quasars caused the reionization.
We selected a redshift range which is lower than is generally considered
for appropriate for the complete reionization of the universe \citep[see
review by][]{LBrev01, Fan01} because before the universe is substantially
ionized, Ly$\alpha$ may be effectively suppressed by resonant scattering
of the Ly$\alpha$ by the neutral intergalactic medium or by the damping
wings of a largely neutral medium.  Thus at or before the epoch of
reionization, there may be no observable Ly$\alpha$ emission with which to
obtain redshifts.  Also, the ionizing photon density necessary to reionize
the intergalactic medium increases as (1+z)$^3$ \citep{Madauradtrans99}
and it is unlikely that the ionizing photon output of the LBGs was the
sources of reionization even if such output remained constant
from z=3 to the reionization at z$>$6 \citep{FDP02}.  Moreover, the time
scale between redshifts of 5.5 and 7 to 10, which is the current best
estimate for when reionization likely occurred \citep[][and references
therein]{LBrev01}, is less than several hundred Myrs for any reasonable
cosmology.  Over such a short time scale, while individual galaxies may
evolve rather dramatically, it is unlikely that the galaxy number density
could evolve very strongly \citep[hierarchical merger models suggest
that it is likely only to be a factor of a few, e.g.,][]{sokasian03,
ciardi03}.  Given these competing constraints, gauging the ionizing
photon density at high redshifts soon after the likely epoch of
reionization may provide a highly significant determination of the
types of galaxies that were responsible for reionization.  Therefore,
for the source of ionization to be galaxies, we seek a population of
galaxies which are still line emitting (i.e., actively forming stars),
with a UV photon density sufficient to keep the intergalactic medium
(IGM) ionized at the observed epoch, and with significant UV continua
suggestive of prolonged star-formation \citep{Starburst99, SUV00} which
is sufficient to ionize the IGM at higher redshifts.  If reionization is
due to QSOs, then we would need to observe many more QSOs than would be
predicted by extrapolating to low luminosity the QSO luminosity function
of \cite{Fan01} for very luminous QSOs (M$_B$$\la-$26).

\section{Observations and Object Selection}

\subsection{Broad-band Observations}

We obtained deep multi-color broad band observations of an approximately
44 arcmin$^2$ field using the imaging spectrograph FORS2 on UT4 of
the VLT. The field was chosen to have minimal extinction from the
maps of \cite{Sextinction98}, low (cirrus) 100$\mu$m emission, to
be south of the latitude of the VLT (to point out of the prevailing
winter wind) and to be easily observable from June through to October.
The data were obtained in ``service mode" on the nights of June 17,
July 6, 7, and 12, 2002. The data were taken as a sequence of dithered
exposures with a net integration time of 2.6 hours in $R-$, 1 hour
each in the $I-$ and $z-$bands.  The images were processed in the
standard way but were flat-fielded using images generated by masking
out all pixels with surface brightnesses between $\pm$3 sigma of the
background noise and then averaged without the images being aligned.
Constructing the flat-field frame in this manner resulted in final
images that are characterized by background levels across the image
that are flat to within much better than 1\% of the average value. The
final combined image had a noise level consistent with that expected from
Poissonian statistics given the sky background and number of individual
exposures. Conditions were photometric throughout the observations.
The final calibration was determined through observations of stars in
the field of PG1323-086 and Mark A \citep{Landolt92} for the $R$ and $I$
filters and by observations of the spectrophotometric standard LTT377
\citep{Hamuy94} for the $z-$band filter.   The final images each has
a total area over which the total integration time contributes to the
signal of 44 square arc minutes.  The 3$\sigma$ detection limits in a 2
arc second diameter circular aperture are $R_{AB}$=27.8, $I_{AB}$=26.7,
and $z_{AB}$=26.0.  The detection limits are consistent with those
expected from the count-rates and sky backgrounds (i.e., the uncertainties
in the photometry do not have significant contributions from errors in
the flat-fielding or any other calibration used in the data reduction).

\subsection{Object Selection for the Multi-Object Spectroscopy}

After the images were completely reduced and the flux calibration
determined, we then made an object catalog using the galaxy photometry
package Sextractor version 2.2.2 \citep{Sextractorref}.  Our final source
catalog was constructed using the I-band image as the detection image with
which to center the apertures and estimate the magnitudes in both the R-
and z-bands.  This method was used as our goal was to find objects
with large spectral breaks between the R- and I-bands, indicating that
they may be at high redshifts.  However, we were concerned that selecting
candidate high redshift galaxies based solely on their detection in
one band might lead to spurious sources.  Therefore, as a check of the
centering of the apertures, the influence of the noise properties of
individual colors or images, etc., we also generated catalogs using
the R-band, z-band, and a simple sum of the 3 images as the detection
images.  Given the color criteria we used to select our target galaxies
for follow-up spectroscopy, we noted no significant dependence on the
image used to select objects other than those given by differences
in the detection limits (like limiting color and source densities).
The final magnitudes adopted for each of the sources is from a 2 arc
second diameter circular aperture based on the aperture position given
using the I-band image for source detection. Although the total imaged
area was approximately 44 arcmin$^2$ we excluded sources near the edge
of the frames to minimize possible flat-fielding errors, differences in
sensitivity due to non-uniform exposure, or sources partially falling
off the imaged region, all of which can affect the photometry.  Moreover,
part of the area was vignetted by the guide probe for some of the image
and this region was also excluded.  Taking these into account, our sample
was selected from the central $\approx$41.3 arcmin$^2$ of the final image.

\begin{figure}
\epsscale{1.0}
\plotone{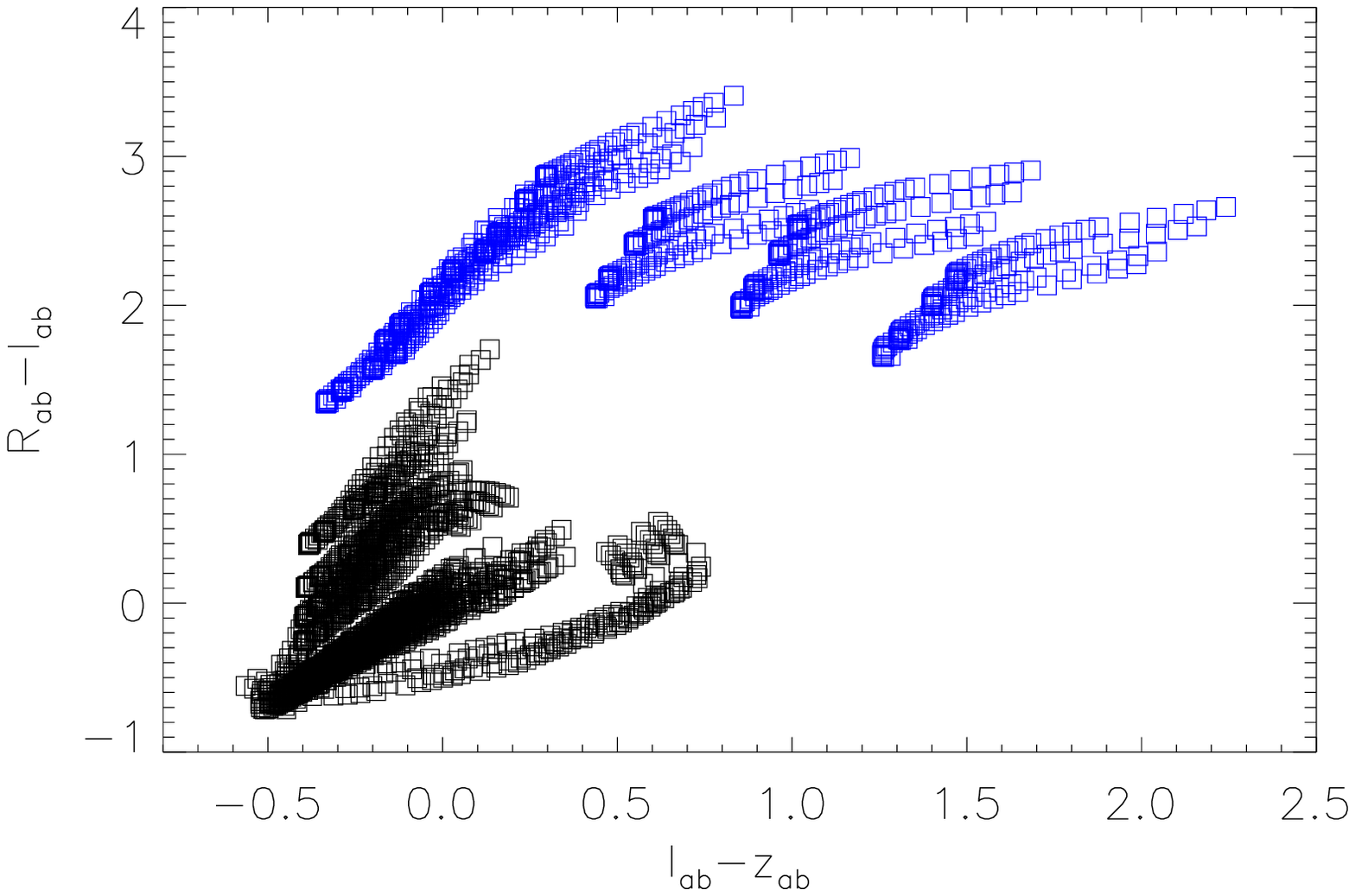}
\caption{The expected locus of colors,R$_{AB}$-I$_{AB}$ versus
I$_{AB}$-z$_{AB}$, for 0.1$<$z$<$5.9.  The models were generated
using the code from \cite{Pegase97}, were all single burst
models with ages that range from 5 Myrs to 14 Gyrs and have
had extinctions at V, A$_V$=0.1, 0.3, 0.7, and 1.0, applied assuming
the \cite{calzetti2000} extinction law.  In addition, we have applied
the absorption of the IGM short-ward of rest-frame Ly$\alpha$ using
the prescription given in \cite{MadauLBGs96} but modified for z$>$4.8
to account for the additional IGM opacity above the \cite{MadauLBGs96}
estimate observed in the SDSS QSOs \citep{Fan02}. For each redshift, all
models with ages younger than the age of the universe at that redshift
are shown. Galaxies with redshifts between 4.8$<$z$<$5.9 galaxies are
shown as blue squares.  As can be seen, galaxies with 4.8$<$z$<$5.9
span a wide range of I$_{AB}$-z$_{AB}$ but almost all are redder than
R$_{AB}$-I$_{AB}$$\ge$1.5, our adopted color selection.
\label{colorcolormodel}}
\end{figure}

\begin{figure}
\epsscale{1.0}
\plotone{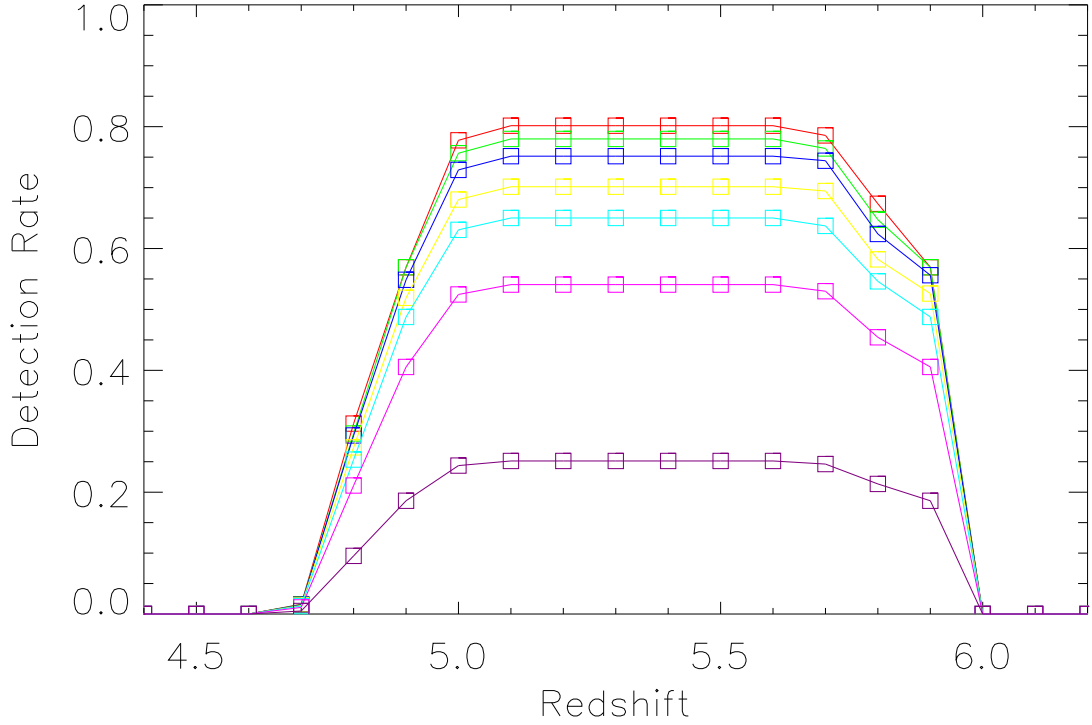}
\caption{The detection rate of the galaxies as a function of magnitude
and redshift.  The magnitudes from top to bottom are for I$_{AB}$=23.25,
23.75, 24.25, 24.75, 25.25, 25.75, and 26.25.  The detection rate for
any magnitude is reasonably uniform between 4.8$<$z$<$5.8.  Due to the
small number of sources with spectroscopic redshifts, we do not plot a
histogram of the redshift distribution but note that the spectroscopic
redshifts are consistent with the estimated detection rate as a function
of redshift.  However, due to the strong night sky emission over the
wavelength range of observed Ly$\alpha$, the redshifts measured correspond
mainly to regions relatively devoid of strong night sky lines.  Thus the
redshift distribution is most strongly influenced by the wavelengths of
the strong night sky and much less so by the selection function of the
objects themselves.
\label{detectionrate}}
\end{figure}

As stated in the introduction, our goal is to use the opacity
of the IGM short-ward of Ly$\alpha$ to select objects with large
magnitude breaks between the $R-$ and $I-$bands.  Using the analysis
of \cite{Madauradtrans99}, \cite{SC02}, and \cite{Fan02} of the
amount of IGM absorption expected short-ward of Ly$\alpha$, along
with spectrophotometric models of early star formation in galaxies
\citep{Pegase97} and existing high redshift quasar spectra, we determined
that a color selection of $R_{AB}-I_{AB}\ge1.5$ will include all
sources at $z>4.8$ (Fig.~\ref{colorcolormodel}).  This selection is
mainly driven by the IGM opacity and thus using an accurate estimate
of it is crucial to selecting galaxies at high redshifts.  Our imaging
data were reliable and relatively complete to $I_{AB}=26.25$ so our
primary target selection was $I_{AB}<26.25$ and $R_{AB}-I_{AB}\ge 1.5$.
These objects could in principle be at any redshift above $z>4.8$, but
in practice are likely to be at $z<5.8$ due to the increasing effect of
IGM opacity on the observed $I-$band at increasing redshift. The IGM
introduces little or no appreciable ``reddening'' in $I-z$ at z=4.8,
increasing to $\sim1.5$ mags of ``reddening'' by $z>5.8$ in these models
(Fig.~\ref{colorcolormodel}). As $R-I$ rapidly increases with redshift
above $z=4.8$, for example at $z=5$, $R-I>2$, the chosen color cut is
generously blue for objects in this redshift range.  Consequently, even
if we have excluded some objects at z$<$5 through our color selection,
the volume that is probed by these observations will be barely effected
(Fig.  \ref{detectionrate}).

Given the number of slit-lets that can be used in the FORS2 Mask
Exchange Unit (see next section), we were able to observe more than
just our primary sample. Consequently, we relaxed our criteria to
$I_{AB}<26.4$ and $R-I>1.0$, including as many objects as feasible,
concentrating on those with the reddest $R-I$ colors.  We expected
few, if any of the sources in this expanded sample to be at $z>4.8$,
essentially only those that had $R-I>1.5$, but with $I_{AB}> 26.25$.
Where there was space on the masks, we included Extremely Red Objects
(EROs) selected from the VLT imaging and a separate $K-$band AAT
IRIS2 image (to be reported on elsewhere).   One object was included
as it was very much brighter in the R-band than in $I$ and undetected
in $z$.  When selecting objects on the basis of color, we did not rely
solely on the Sextractor cataloged $R-$band magnitudes, we examined the
$R-$band images of all objects with $I_{AB}<26.4$ and $R_{AB}>27.2$ by
eye. This allowed us to include objects where the $R-$band magnitudes
were overestimated by Sextractor due to crowding with other objects or
from unreliable background estimation. We also examined sources detected
by Sextractor in the $I-$band alone, rejecting those where the source
appeared bogus. This only affected a few of the faintest objects.

The $z-$band was used to constrain the observational priority in an
attempt to remove intrinsically red objects (cool stars and sub-stellar
objects), but again, there was sufficient slit-let numbers that this
criterion was not strictly observed and such objects were included in
spectroscopy to confirm their photometric classification.  In the rest
of this paper, we will only discuss the objects that strictly meet the
primary photometric criteria.

\subsection{Multi-Object Spectroscopy}

Spectra of these sources were obtained using the Mask Exchange Unit
(MXU) in FORS2 on UT4. Key to the success of these observations was
the use of new red-enhanced MIT CCD chips, which benefited from high
Quantum Efficiency at red wavelengths (70 per cent at 9000 \AA) and
extremely low fringe amplitudes. The observations were carried out
in service mode on the nights of August 6 and 8, 2002.  We observed
with the 300 l mm$^{-1}$ grism blazed in the I-band and an OG590 order
separating filter.  The final spectra span the range of about 6000\AA \
to 1.1$\mu$m for slits placed near the center of the field. The set-up
included two masks each of which were observed partially on those
two nights.  The total integration time per mask was 4 1/3 hours.
Generally, each slitlet was 10" in total length and 1" in width.
The data were reduced by bias subtracting frames with zero exposure time
and no illumination, flat-fielded using normalized spectra of a continuum
source, and wavelength calibrated using observations through each mask of
a comparison lamp.  Each mask had 20 sets of exposures which were dithered
along the slit.  After bias subtraction and flat-fielding, the sky was
subtracted using a average combination of 3 frames taken closest in time
and then had a first order polynomial subtracted from each column after
the first pass at sky subtraction.  The separate spectra for each slit
were then aligned and averaged together.  The spectra were then rectified
to wavelength calibrate them and then flux calibrated using the standard
spectrophotometric standard LTT 7987. The excellent cosmetic quality of
the chips in FORS2 along with their very high quantum efficiency and lack
of fringing in the red meant that we could often spectroscopically detect
very faint objects ($z_{AB} \ge 25$) to 1 micron (Fig.~\ref{twodspectra}).

\begin{deluxetable}{lcccccccc}
\tabletypesize{\scriptsize}
\tablecaption{Properties of the Primary Sample\label{tbl-1}}
\tablewidth{0pt}
\tablehead{
\colhead{Designation}&\colhead{R$_{AB}$}& \colhead{I$_{AB}$}&
\colhead{z$_{AB}$}&\colhead{$\lambda_{line}$}&\colhead{log f$_{line}$}&
\colhead{Redshift}& \colhead{log L$_{line}$}& \colhead{M$_{AB}$(1700\AA)}\\
\colhead{(1)}&\colhead{(2)}&\colhead{(3)}&
\colhead{(4)}&\colhead{(5)}&\colhead{(6)}&
\colhead{(7)}&\colhead{(8)}&\colhead{(9)}
}
\startdata 
BDF1:9&$>$27.8&26.2&25.8&...&...&...&...&$-$20.3 \\
BDF1:10&$>$27.8&26.0&25.3&8191.8&$-$16.61&5.7441&42.94&$-$21.3 \\
BDF1:11&$>$27.8&25.8&25.5&7078.2&$-$17.80&4.8223&41.58&$-$20.8 \\
BDF1:14&$>$27.8&26.0&24.9&...&...&...&...&$-$21.7 \\
BDF1:18&$>$27.8&26.0&25.4&7315.5&$-$17.62&5.0175&41.80&$-$21.0 \\
BDF1:19&$>$27.8&26.0&25.4&8351.4&$-$17.51&5.8696&42.07&$-$21.1 \\
BDF1:26&$>$27.8&25.6&25.9&7362.0&$-$17.46&5.0558&41.97&$-$20.7 \\
BDF2:12&$>$27.8&26.2&$>$26.0&...&...&...&...&$>-$20.5 \\
BDF2:13&$>$27.8&26.0&$>$26.0&...&...&...&...&$>-$20.5 \\
BDF2:15&$>$27.8&25.9&$>$26.0&...&...&...&...&$>-$20.5 \\
BDF2:17&$>$27.8&26.2&$>$26.0&...&...&...&...&$>-$20.5 \\
BDF2:19&$>$27.8&26.1&25.2&8083.0&$-$16.60&5.6488&42.94&$-$21.4 \\
I1020&$>$27.8&26.1&$>$26.0&...&...&...&...&$>-$20.5 \\
\enddata
\tablecomments{Col. (1) -- Source name. Cols. (2-4) -- Magnitudes
in the AB system. Col. (5) -- Wavelength of identified line in
\AA. Col. (6) -- Logarithm of the integrated flux of the line in ergs
s$^{-1}$ cm$^{-2}$. Col. (7) -- Redshift assuming the observed line
is Ly$\alpha$. Col. (8) -- Logarithm of the total line luminosity
assuming H$_0$=70 km s$^{-1}$ Mpc$^{-1}$, $\Omega_{matter}$=0.3,
$\Lambda$=0.7. Col. (9) -- Absolute AB magnitude at 1700\AA,  calculated
from the $z-$band magnitude assuming no color term.  The sources without
measured redshifts are assumed to be at z=5.3 which is approximately at the
peak selection efficiency (see Fig.~\ref{detectionrate}.}

\end{deluxetable}

\begin{deluxetable}{lccccccc}
\tabletypesize{\scriptsize} \tablecaption{Other galaxies at $z>4$ in
the Expanded or Serendipitous Samples\label{tbl-2}} \tablewidth{0pt}
\tablehead{ \colhead{Designation}&\colhead{R$_{AB}$}&
\colhead{I$_{AB}$}&
\colhead{z$_{AB}$}&\colhead{$\lambda_{line}$}&\colhead{log
f$_{line}$}& \colhead{Redshift}& \colhead{log L$_{line}$}\\
\colhead{(1)}&\colhead{(2)}&\colhead{(3)}&
\colhead{(4)}&\colhead{(5)}&\colhead{(6)}& \colhead{(7)}&\colhead{(8)}
} 
\startdata
BDF1:5 &27.7&26.3&25.8&7142.5&$-$17.17&4.8752&42.22 \\
BDF2:5  &$>$27.8&$>$26.7&$>$26.0&9229.8&$-$17.32&6.5926&42.38 \\
BDF1:19a&$>$27.8&$>$26.7&$>$26.0&6640.9&$-$16.97&4.4626&42.33 \\
BDF1:32a&25.2&24.4&24.3&...&...&4.49&... \\
BDF1:32b&$\sim$25.5&$>$25.7&$>$26.0&6677.5&$-$17.89&4.4927&41.42 \\
\enddata
\tablecomments{Col. (1) -- Source name. Cols. (2-4) -- Magnitudes in the
AB system. For  BDF1:32b Both the cataloged $R$ and $I$ magnitudes
were affected by poor background estimation, due to the presence
of a bright nearby star, hence the limit for the $I-$band magnitude.
Col. (5) -- Wavelength of identified line in \AA. Col. (6) -- Logarithm
of the integrated flux of the line in ergs s$^{-1}$ cm$^{-2}$. Col. (7)
-- Redshift assuming the observed line is Ly$\alpha$. For BDF1:32a,
the redshift estimate is based on the continuum break assuming it is
due to Ly$\alpha$ absorption from the IGM. Col. (8) -- Logarithm of
the total line luminosity assuming H$_0$=70 km s$^{-1}$ Mpc$^{-1}$,
$\Omega_{matter}$=0.3, $\Omega_\Lambda$=0.7.}

\end{deluxetable}

\section{Results}

\subsection{Results from the Spectroscopy}\label{specresults}

There were 18 objects which met our strict primary selection criteria
of $I_{AB}<$26.25, $R_{AB}-I_{AB}\ge1.5$. These split neatly into
two groups based on $I-$band magnitude. The first group of five had
$I_{AB}<24.5$. Three were point sources with the colors of main sequence
stars and one had the colors and morphology of a $z\sim 1$ elliptical,
with its $R_{AB}-I_{AB}$ color just making the color cut. The fifth object
was stellar in appearance but had $R_{AB}-I_{AB}>3$ and $I_{AB}-z_{AB}\sim
0.6$ making it a candidate quasar. All except one of the stellar objects,
including the quasar candidate, were spectroscopically observed. None
were found to be at $z>4.8$. The candidate elliptical was also confirmed
spectroscopically.

\begin{figure}
\epsscale{0.8}
\plotone{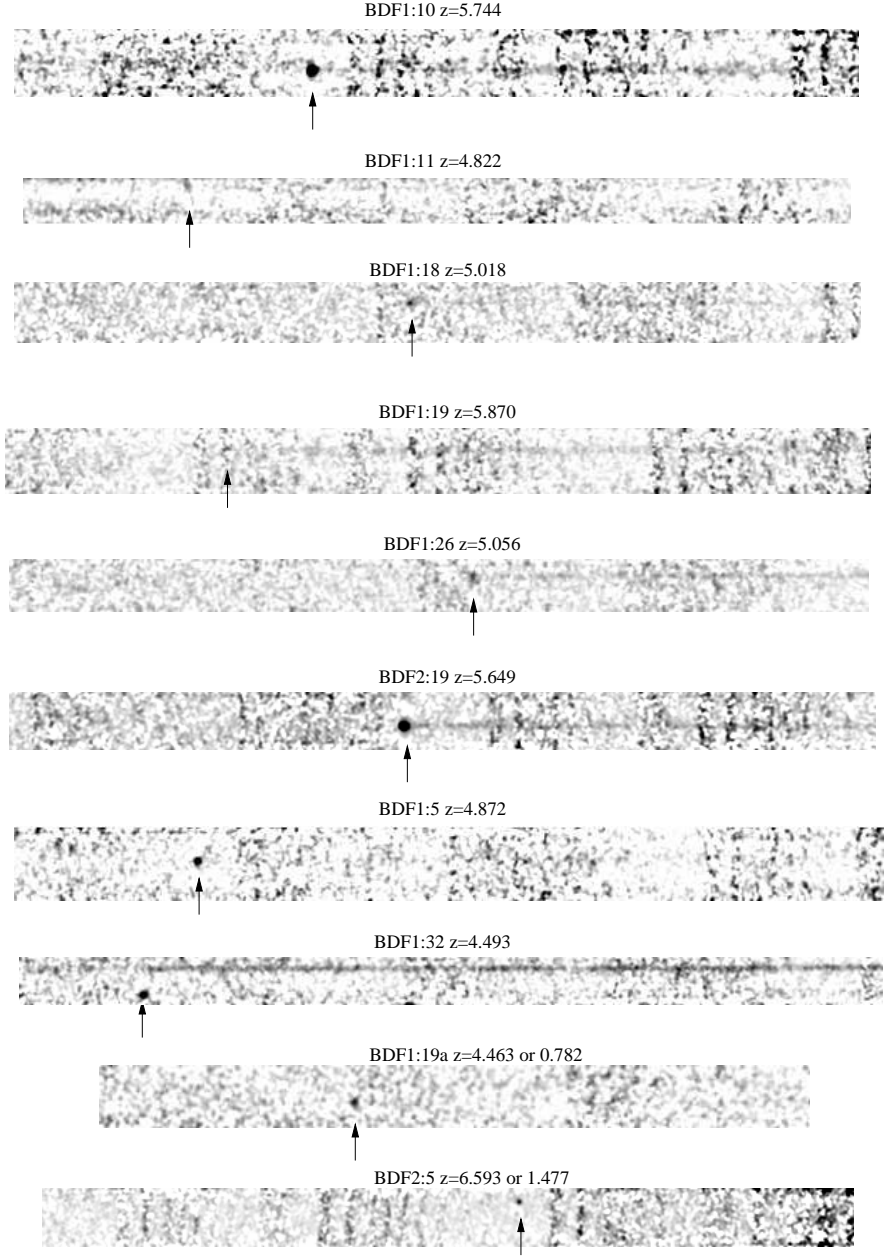}
\caption{Two-dimensional spectra of the galaxies for which we have
estimated the redshifts.  From top to bottom: BDF1:10 (z=5.744), BDF1:11
(z=4.822), BDF1:18 (z=5.018),  BDF1:19 (z=5.870), BDF1:26 (z=5.056),
BDF2:19 (z=5.649), BDF1:5 (z=4.872), BDF1:32 (both BDF1:32a and b have
z$\approx$4.49), BDF1:19a (z=4.46; the blue end of the slit-let spectrum
of BDF1:19 shown near the top of the Figure), and BDF2:5 (z=6.59).
The bottom four galaxies are not part of the primary sample of LBGs.
Each of the displayed spectra is approximately 1500\AA \ in length.
\label{twodspectra}}
\end{figure}

\begin{figure}
\plottwo{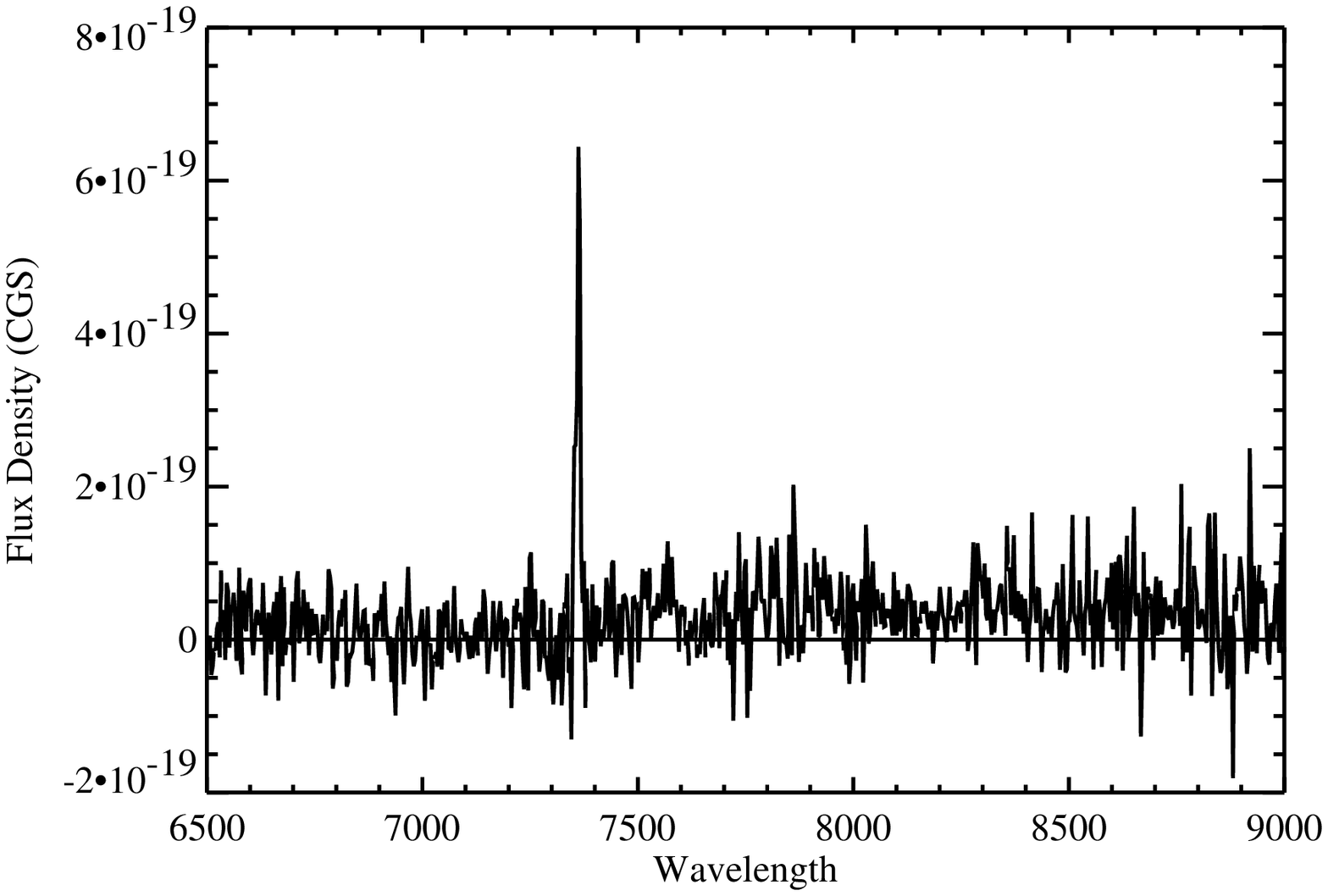}{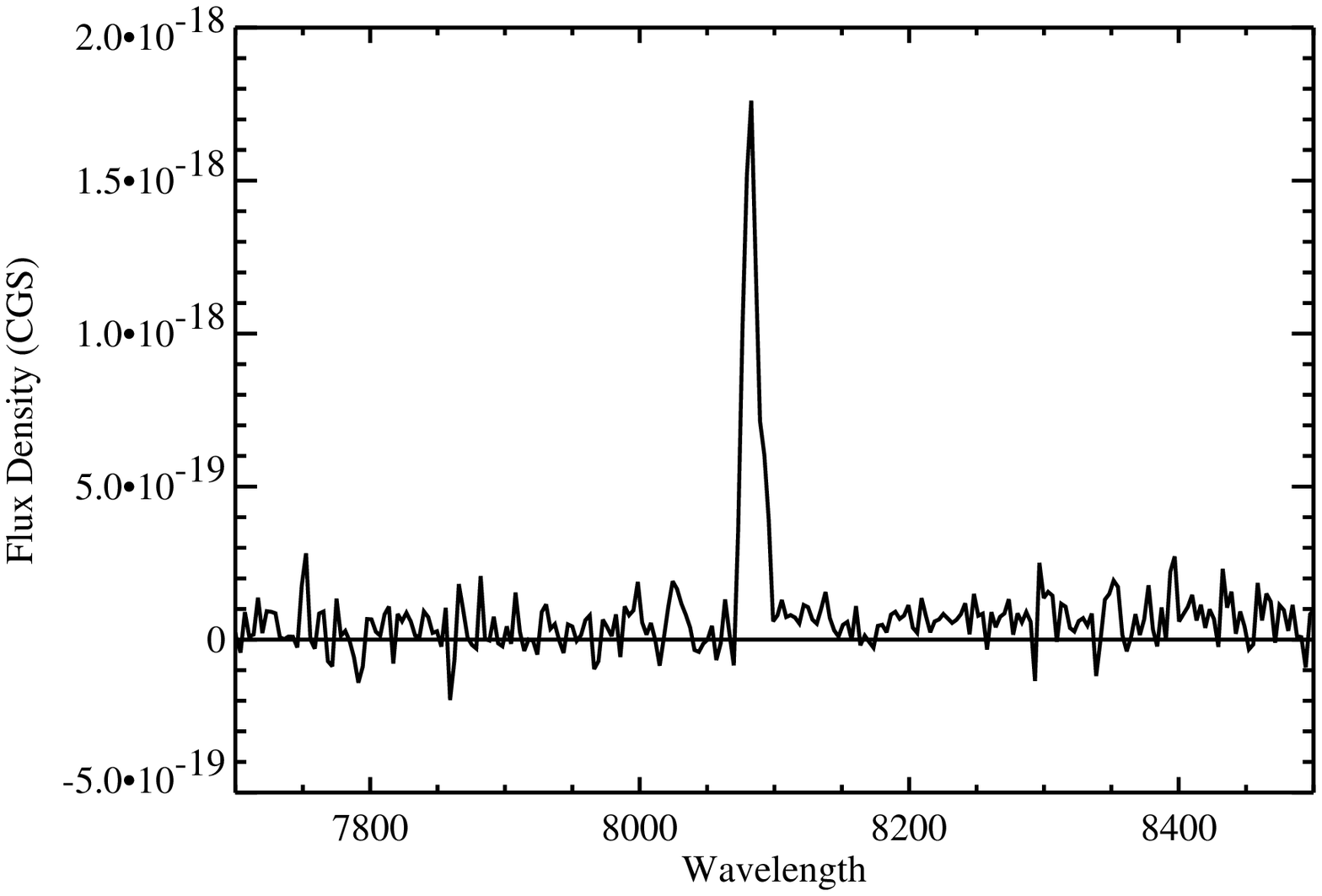}
\caption{One dimensional spectra of the spectroscopically confirmed
LBGs selected to have $(R_{AB}-I_{AB})>$1.5, I$_{AB}$$<$26.1, and
$R_{AB}$$>$27.6.  {\it (top)} BDF1:10 and BDF1:11.  {\it (middle)}
BDF1:18 and BDF1:19.  {\it(bottom)} BDF1:26 and BDF2:19.  The large
continuum break between R and I in all of the sources and the obvious
line asymmetry in the strongest line emitters imply that the detected
lines are likely Ly$\alpha$ emission at high redshifts.
\label{onedspectra}}
\end{figure}

The second group of 13 sources all had $25.5<I_{AB}<26.25$ and
$R_{AB}>27.8$.  All were undetected in $R$ and half undetected in $z$.
All 13 sources are listed in Table 1 along with their redshifts when
measured. We obtained spectra for 12 of these sources, the other source
was excluded due to slit contention (by which we mean that the objects
were aligned such that it was impossible to place two slits that would not
overlap spatially and would still be long enough to provide for accurate
sky subtraction). Of these sources, 6 have spectroscopically-determined
redshifts and their spectra are shown in Figs.~\ref{twodspectra} and
\ref{onedspectra}.  Our sample selection was optimized to detect objects
with redshifts between z=4.8 to 5.8.  To determine this selection,
we carried out an analysis similar to that presented in \cite{SLBGs4}.
In addition to knowing the expected colors of high redshift galaxies,
the completeness of the observations need to be determined.  We made the
incompleteness estimate in two ways.  First, we distributed galaxies of
different magnitudes and colors (effectively redshifts) randomly across
the frame, then attempted to recover them with the same procedure and
color selection used when making the original galaxy catalogs. This
analysis was similar to that performed in \cite{SLBGs4}, the only
difference was that we used the actual galaxies detected with known
redshifts and colors instead of model galaxies.  Second, to estimate
the relative influence of the possible sources of incompleteness (i.e.,
source confusion, incompleteness as a function of magnitude, and the
influence of our color selection criteria on the completeness as a
function of redshift) we determined the incompleteness in a three-step
process.  We determined the effective area of our images by recovering
objects placed at random across the I-band frame. A range of magnitudes
were used in order to determine the fraction of galaxies recovered.
We then determined the incompleteness directly by determining the
deviation of the faint galaxy surface density from a power-law fit
to the source surface density as a function of magnitude over the
range of magnitudes where the data were complete (in the I-band,
this was 22.5$<$I$<$25.5).  We also estimated the incompleteness as a
function of redshift due to our color-selection criteria.  For this, we
carried out a Monte Carlo simulation using the range of model parameters
used in Fig.~\ref{colorcolormodel}, distributing these randomly over
the redshift range 2 to 6.5.  The results of the Steidel-like and the
three-step-process estimates agree within statistical uncertainties.  The
final results of the second method are shown in Fig.~\ref{detectionrate}.
Given the range of confirmed redshifts, we have succeeded as expected.

Spectra of several of the other six sources showed evidence for a Lyman
break in that redshift range, but none as sharp as the line emitters
due to limited signal-to-noise. Consequently we do not assign these
an exact redshift. Others were too faint to reliably detect spectral
features in the continuum. None showed features that indicated redshifts
of $z<4.8$. Also, the redshift distribution of the line-emitters is
not uniform between z=4.8 and 5.8 but in fact fall mainly near the low
and high ends of the redshift range.  This is most likely due to the
wavelength distribution of the strong night sky lines, since the redshift
measured tend to be in regions devoid of strong sky lines.  Therefore it
is possible that some of the sources without spectroscopically-determined
redshifts are such because their redshifted Ly$\alpha$ line falls in
a region of strong night sky line emission. In what follows we refer
to these 13 objects as the primary sample.  The inclusion of the last
object is justified in that it was only excluded from the spectroscopic
sample due to a conflict in slit position with another candidate high
redshift source.

The strong line-emitters in this sample of 13 are characterized by
extremely high rest-frame equivalent widths, more than 30-50\AA \
in the rest-frame.  The Ly$\alpha$ line luminosities implied by line
fluxes are about 4 $\times$ 10$^{41}$ to almost 10$^{43}$ ergs s$^{-1}$.
Using the models of \cite{Starburst99} and \cite{CF93} and assuming case B
recombination to convert Ly$\alpha$ to H$\beta$ line luminosity, we find
that UV continuum and line emission is indicative of an extremely young
galaxy (ages$<$10$^{7-8}$ years given the uncertainty in the equivalent
width) and star-formation rates of up to about 20 M$\sun$ yr$^{-1}$
(neither of these estimates take into account possible extinction).

None of the objects in the primary sample have
spectroscopically-determined redshifts below z$<$4.8.  Only one object
in the expanded sample was found to be at $z>4.8$. This had an $I-$band
magnitude of $I_{AB}=26.3$, just too faint to be included in the primary
sample, and was undetected in $R$, so had $R_{AB}-I_{AB}\ge1.5$. It
has a redshift of $z=4.88$. This indicates that selecting objects with
$I_{AB}>25$ and $R_{AB}-I_{AB}\ge1.5$ is a reliable and efficient way
of selecting galaxies at $z>4.8$ and out to at least $z=5.8$, given the
results of our spectroscopy.

One other object in the expanded sample, BDF1:32b, was found to be
at $z>4$ (Table 2).  This was detected in both $R$ and $I$ and had a
relatively blue $R-I$ color. However, we note that BDF1:32 lies near a
bright star which affected the photometry significantly.  Practically
all of the R-band light came from a strong emission line at 6675 \AA
(Fig.~\ref{twodspectra}). This is taken to be Ly$\alpha$ at $z=4.46$.
Another continuum-only object in the slit 5 arc sec north of the
line-emitter has a sharp break in the continuum at the same wavelength,
with little or no continuum flux short-ward of $\approx$ 6675 \AA
(BDF1:32a). These two objects are therefore likely to be at the same
redshift, one a strong line emitter, the other a continuum-only Lyman
break galaxy (Fig.~\ref{twodspectra}).

\subsection{Surface Densities of High Redshift LBGs}


How do the luminosities and surface densities of our sources compare to
those of the Lyman break galaxies in \cite{SLBGs4} at $z=3$ and 4?  Given
the small area of our survey and the small number of sources, we do not
attempt to determine a luminosity function for the sources. Determining
the luminosity function is complicated by the diminution of the $I-$band
flux from an object as the redshift increases from z=4.8 to 5.8 due to the
opacity of the IGM. Instead, we constructed a simulation of the expected
number counts in $I_{AB}$ over this redshift range assuming no evolution
in the luminosity function taken from \cite{SLBGs4} for z$\approx$3.0 and
4.1 LBGs (converted to the cosmology H$_0$=70 km s$^{-1}$ Mpc$^{-1}$,
$\Omega_{matter}$=0.3, $\Omega_\Lambda$=0.7) and that the galaxies are
distributed uniformally in the {\it completeness-corrected} volume (i.e.,
the completeness correction has been applied as a function of magnitude
and redshift; Fig.~\ref{detectionrate}).  We assume that the absolute
magnitude at 1700\AA~ relates directly to the observed $z-$ band magnitude
with no color term. We modelled the $I-z$ colors of young galaxies (from
the models of \citeauthor*{Pegase97}) over this redshift range including
the effects of IGM opacity \citep[from][modified using the determinations
from \cite{Fan02}]{Madauradtrans99}. An approximate fit to the color
of these objects is given by $I_{AB}-z_{AB}=1.5\times (z-4.8)$ over the
redshift range $4.8<z<5.8$.  With these assumptions we distributed the
galaxies assuming their magnitudes followed the luminosity function from
\cite{SLBGs4}, assigned a redshift with the appropriate difference in
the distance modulus between z=4.1 and the assigned redshift, and then
applied the diminution due to the IGM.  The incompleteness correction
was included using the estimates outlined in \S \ref{specresults}.

\begin{figure}
\plotone{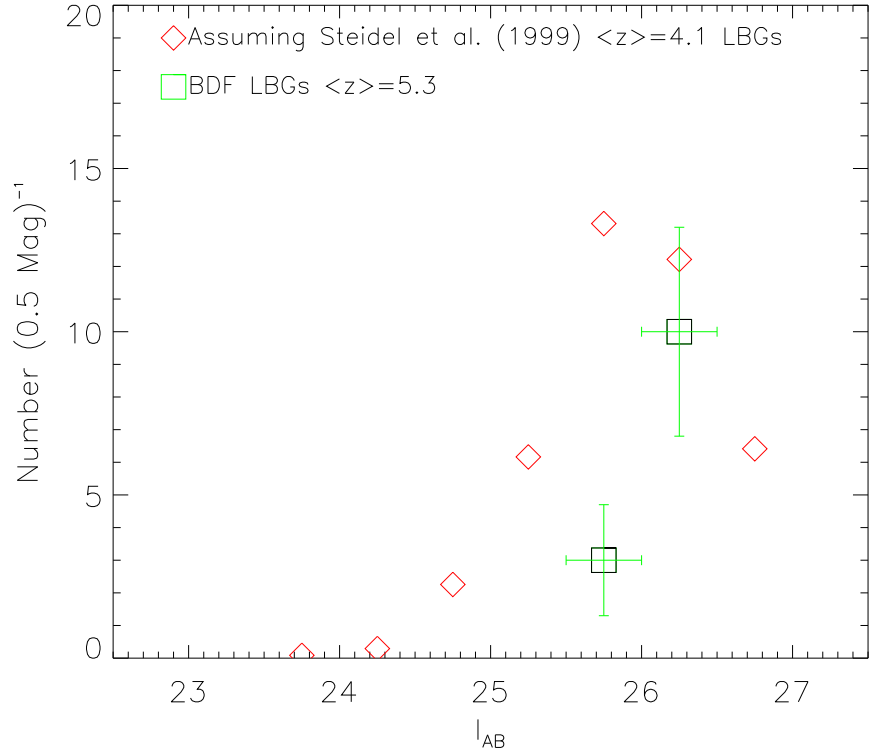}
\caption{A comparison of the predicted number of LBGs at z$\approx$5.3
and the observed number in our field.  To construct this plot, we made
a simple monte-carlo simulation of the expected number of LBGs assuming
that the co-moving volume density of LBGs at z$\approx$3.0 and 4.1 LBGs
from \cite{SLBGs4} remained unchanged to z$\approx$5.3, used a simple
scaling for the effect of the absorption due to the IGM short-ward of
Ly$\alpha$ on the $I-z$ color, that the $z-$band magnitude is related
magnitude of the LBGs at z$\approx$4.1 through the difference in the
distance modulus, and that the redshift distribution of the sources was
similar to that modelled in Fig. \ref{detectionrate} between 4.8 and 5.8.
In addition, we applied the incompleteness of our data to the simulated
distribution.  The results of the simulation are shown as red diamonds.
The green squares represent our observed number of sources in the field.
The uncertainties of the observed number of galaxies are Poissonian.
\label{LBGcomparison}}
\end{figure}

The simulated number counts, including the incompleteness as a function
of magnitude and redshift are shown in Figure~\ref{LBGcomparison}. Above
$I_{AB}<25.5$ we have no candidate galaxies at $z>4.8$, where as of-order
10 are predicted.  Down to $I_{AB}<26.25$, about 40 sources are predicted
compared to thirteen detected in our primary sample. Although the number
of detected sources are below those expected if the luminosity function
of UV bright Lyman break galaxies does not evolve from $z=3$ to $z>5$,
the differences are not highly significant except at the brightest end.
Even then the cosmic variance might account for much of this difference
given the relatively small area of our field.  But taken at face value,
our results suggest that the luminosity function at these high results has
shifted to fainter magnitudes and is steeper at the bright end compared
to the luminosity function at z=4.1.

\begin{figure}
\plotone{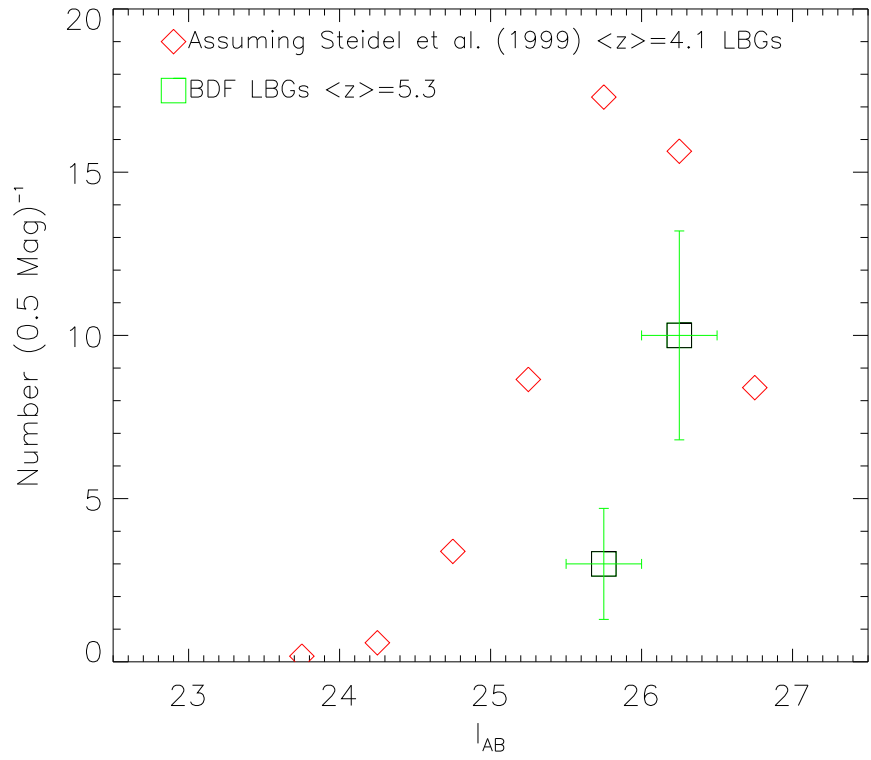}
\caption{This is the same similation as presented in Fig. \ref{LBGcomparison}
except now we have assumed $I_{AB}-z_{AB}=1.0\times (z-4.8)$.
\label{LBGcomparison2}}
\end{figure}

The estimate of any significant under-density of sources relies
predominately on the relationship between $I-z$ and redshift. The range of
$I_{AB}-z_{AB}$ colors seen in the models in Fig. \ref{colorcolormodel}
results from the strong absorption due to the IGM with a smaller
contribution from range of extinction and ages.  The span in color due
to redshift and IGM absorption is about 2 magnitudes, while that due to
reddening and age is about 0.5 magnitudes for any one redshift. The range
of $I_{AB}-z_{AB}$ colors observed in the galaxies of our primary sample
at $<$0.2 to 1.1.  Although the number of spectroscopically-determined
redshifts is small, the galaxies at z=5.6 to 5.8 do have the reddest $I-z$
colors of our sample ($I_{AB}-z_{AB}$ = 0.6 or 0.7), consistent with the
dominant influence of the attentuation due to the IGM.  These colors
are consistent with little reddening and young ages ($\la$10$^8$ yrs)
mainly due to the strong attenuation of the IGM at the highest redshift.
And since dust destroys Ly$\alpha$ preferentially and only galaxies with
young ages have strong Ly$\alpha$ emission \citep[e.g.,][]{Starburst99,
CF93}, such a conclusion is also supported by the high rest-frame
equivalent widths observed in Ly$\alpha$.  Given the dominant influence
of the IGM opacity on the colors of the galaxies, and the lack of
extremely red $I-z$ colors, one might be concerned about how the choice
of $I_{AB}-z_{AB}=1.5\times (z-4.8)$ used in our simulation might
bias our conclusions.  To gauge the influence the relationship between
color and redshift on the predicted numbers of LBGs, we ran the surface
density simulation with a variety of coefficients for linearly relating
redshift and $I_{AB}-z_{AB}$ (e.g., $I_{AB}-z_{AB}=1.0\times (z-4.8)$
and $I_{AB}-z_{AB}=-0.5 + 1.0\times (z-4.8)$) that would be consistent
with both the model and observed galaxy colors as a function of redshift.
The results imply that our assumption of $I_{AB}-z_{AB}=1.5\times (z-4.8)$
is conservative.  For example, adopting $I_{AB}-z_{AB}=1.0\times (z-4.8)$
(which is more in line with the observed range of colors and redshifts
than is $I_{AB}-z_{AB}=1.5\times (z-4.8)$), about 50 sources are predicted
compared to thirteen detected in our primary sample and the discrepancy
at the faint magnitudes becomes significant (Fig. \ref{LBGcomparison2}).
Therefore, any relationship between color and redshift with a weaker
dependence on redshift or which reaches less extreme $I_{AB}-z_{AB}$
colors at z=5.8 than we have assumed, tends to increase the significance
of the observed under-density of sources compared to LBGs at lower
redshifts.

The justification for the analysis just presented is that the IGM has
a substantial impact on the resulting magnitudes and number densities
of the sources at these redshifts.  Therefore, it is imperative that
its affect be accounted for properly.  In spite of this caveat, to
further our comparison with LBGs at low redshifts, we estimated the
total co-moving densities and UV luminosity density from sources with
4.8$<$z$<$5.8.  Applying the incompleteness corrections, we find the
total co-moving density of sources with $M_{AB}(1700\AA)\ga-20$ of about
10$^{-3.3}$ Mpc$^{-3}$.  The estimated UV luminosity density is about
log $\rho_{UV}$$\approx$25.7 ergs s$^{-1}$ Hz$^{-1}$ Mpc$^{-3}$.  Both of
these estimates are approximately a factor of 3 below that estimated at
z$\approx$4 \citep{SLBGs4}.  Although the correction for IGM opacity
is uncertain and that both the total co-moving space densities and UV
luminosity density strongly reflect the number densities of sources
at the faintest magnitudes where the relative under-density is small,
these estimates suggest an overall decline in the number and emissivity
of sources from z$\approx$4 to z$\approx$5.3 consistent with our simulations.

A possible cause of the discrepancy in the expected number of sources at
z$\approx$5.3 and z$\approx$4 could be the effects of surface brightness
dimming which results in systematically underestimating the magnitudes
of galaxies at high redshift.  To provide an estimate of the size of the
effect, the relative difference in surface brightness between z=4.1 and
5.3 is about 1 magnitude arcsec$^{-2}$.  However, surface brightness
dimming is unlikely to have a substantial impact for several reasons.
The images used to select the galaxies for this survey are about one
magnitudes deeper than the ground-based images used in for example
\cite{SLBGs4} thus compensating somewhat for the differences in
surface brightness of the two galaxy samples.  In addition, we note
that none of the objects is resolved significantly in the 0.7 arcsecond
seeing of the images.  In an analysis of similarly selected galaxies
from the Chandra Deep Field South with deep HST ACS images, galaxies
have a half-light radii that span from unresolved at HST resolution
to less than a few tenths of an arcsec \citep[Bremer et al. 2003, in
preparation;][]{stanway03}.  Moreover, if all of the galaxies have a
similar morphology regardless of their magnitude, one would expect the
fainter sources to have their magnitudes significantly underestimated
compared to the brighter sources.  This is because brighter galaxies
will be detected over a broader range of surface brightness than that
of fainter galaxies.  Our results suggest that the difference in the
source densities is significant for the bright high redshift galaxies and
much less so for the faint sources.  Thus, surface brightness dimming
is unlikely to play a significant role in explaining the paucity of
bright sources.

We also investigated the constraints that these results put on the QSO
luminosity function at these high redshifts.  We detected no broad line
sources (QSOs) in our spectra.  Assuming that the redshift coverage is
between 4.8 and 5.8, and applying the completeness correction discussed
above, we find that we find that the co-moving density of QSOs must be
less than about 10$^{-4.8}$ Mpc$^{-3}$ for $-$23.5$\la$M$_B$$\la$$-$20.5
(due to the completeness corrections, the exact density depends on the
magnitude; Fig. \ref{QSOnumlimit}).  The B-band magnitude was estimated
assuming a power-law slope for the QSO spectrum of -0.5 \citep{SSG95,
Fan01}.  This limit on the co-moving space density implies that the
QSO luminosity function as observed by \cite{SSG95} and \cite{Fan01}
while already flatter than the lower redshift luminosity functions,
must turnover at M$_B$$\la-$23.

Several other galaxies at these redshifts have been discovered in
previous studies, but the diversity in galaxy selection methods make
the studies difficult to compare to ours. Those selected by lensing
\citep{Ellis01} or narrow-band imaging \citep{Hu02, Rhoads02} have
continuum magnitudes that are generally below our $I-$band limits. Others
were serendipitously discovered \citep{Dey98}, or based on optical/IR
colors \citep{Ray98}. Techniques comparable to ours were used by
\cite{Dey98} and \cite{Spinrad98} to detect z=5.3 galaxies in individual
Keck and HST images. The pair of galaxies detected by Spinrad et al. in
the HDF-North would have been seen in our ground-based imaging as an
individual source. Both this and the galaxy detected by \citeauthor{Dey98}
would have been discovered by us had they fallen within our field. Given
the relative field sizes and exposure depths, the discovery of these
objects is entirely consistent with the surface density of galaxies at
$z>4.8$ in this study.

Wide-field narrow-band surveys for line-emitting galaxies \citep[e.g.,][]
{RM01, O02} are efficient at detecting the most extreme line-emitters
at high redshift.  None of the galaxies in \cite{Rhoads02} have
sufficient line flux alone to be detected by us in the broad-band
filters given our flux limit. For example, a flux of $5\times 10^{-17}$
ergs$^{-1}$ s$^{-1}$ cm$^{-2}$, is below our detection limit in
both I and z assuming that a majority of the flux from the objects
in these bands arises from a Ly$\alpha$ line.  Adopting the figures
in \cite{Rhoads02}, assuming a $3-\sigma$ I-band continuum detection
limit of $I_{AB}>25.9$, we would expect about 4 objects (taking into
account the incompleteness) with line fluxes of $>1.5\times 10^{-17}$
ergs$^{-1}$ s$^{-1}$ cm$^{-2}$ and $I_{AB}>25.9$ in our central field
of 41.3 arcmin$^2$.  The \citeauthor{Rhoads02} survey is sensitive
to the most extreme line-emitters, our continuum-detected sources in
the primary sample typically have lower Ly$\alpha$ equivalent widths
(although we note that several of them, including the strongest
of the line emitters have equivalent widths $>$100\AA).  Formally,
we detect 2 line emitters with line fluxes greater than $>1.5\times
10^{-17}$ ergs$^{-1}$ s$^{-1}$ cm$^{-2}$ and $I_{AB}>25.9$.  However,
these galaxies both are I$_{AB}$$\approx$26, barely below the $3\sigma$
continuum detection limit of \citeauthor{Rhoads02}  Thus while consistent
within the uncertainties, it is likely that there are more objects with
fainter continuum magnitudes similar to the \citeauthor{Rhoads02} sources,
but with lower line fluxes in our field. If we assume that the Ly$\alpha$
luminosity density measures the unobscured star formation density, it is
therefore likely that a significant fraction (probably most) of the star
formation occurring in our field at $z>4.8$ happens in objects fainter
than our magnitude limit.  If so, the UV luminosity density produced by
these objects is also likely to be more than that produced by objects
above our magnitude limit.

\subsection{Serendipitously Detected Line Emitters: The Highest Redshift
Galaxy?}

Along with objects that we deliberately targeted, we also detected two
potentially high redshift line-only sources serendipitously in our
slit-lets. The first was a line detected at 6640.9\AA~ in the same
slit-let as BDF1:19 (Figure~\ref{twodspectra}; Table 2). The second
was found in a slit-let placed to fill up empty area on the slit
mask (BDF2:5). The line is at an observed wavelength of 9229.8\AA \
(Figure~\ref{twodspectra}; Table 2) which, assuming the detected line
is redshifted Ly$\alpha$, would make it the highest redshift galaxy
known \cite[see][]{Hu02}, at $z=6.6$.  However, the lack of discernible
continuum emission makes this identification uncertain.  The line is
too weak to have a significant asymmetry in our data. It could be
a low redshift [OII]$\lambda$3727 emitter at z=1.4765.  If so, the
upper limit of $z_{AB}$=26.0 implies that it would have a rest-frame
U$_{AB}$$<$$-$17.5 and a rest-frame equivalent width $>>$50\AA.
Both numbers are extreme, but not completely out of the question.
\cite{Tr99} have shown that in the local universe, while such extreme
equivalent widths are generally rare ($<<$10\% of galaxies are so
extreme), high equivalent widths become more common in low luminosity
galaxies. Although the identification of the line as Ly$\alpha$ is not
certain, we note that the redshift is no less secure than other line-only
high redshift galaxy candidates in the literature, or those with very
marginal single-band continuum detections.

\section{Discussion}

Recently, \cite{FDP02} have used the analysis of \cite{S01} and
\cite{papovich01} to estimate the possible contribution of LBGs at
z$\approx$4 to the reionization of the Universe.  They argue that
for reasonable assumptions, the relatively young age and insufficient
co-moving space density, LBGs at z$\approx$4 are unlikely to be the
progeny of the sources that reionized the Universe.  They also show
that if the luminosity function of the LBGs does not evolve between
the reionization epoch and $z=3$, then they cannot provide the bulk of
the UV photons that reionize the universe. The higher the redshift of
reionization, the larger the deficit.

We find that there may be a decline in the number of luminous star-forming
galaxies from z$\approx$3.0 and 4.1 \citep{SLBGs4} to z$\approx$5.3.
The number of sources we observe at I$_{AB}<25.5$ is less than expected
(zero observed and about 10 predicted) for a constant co-moving density
of sources over that redshift interval.  Our calculated UV photon density
for our primary sample over the redshift range $4.8<z<5.8$ falls
below that of $z\sim 3$ LBGs, insufficient to ionize the IGM in the
observed volume. Another source of UV photons is required. Moreover,
these sources are observed within only a few hundred million years after
reionization (assuming this happens at $z<7-10$). For galaxies such as
those in our primary sample to provide the bulk of reionizing photons,
a dramatic and unlikely decrease in their co-moving density would have
to occur over this short timescale.  For this deficit to be accounted for
by cosmic variance, the field would have to be considerably more than an
order of magnitude under-luminous relative to the average field at $z>5$.

\begin{figure}
\plotone{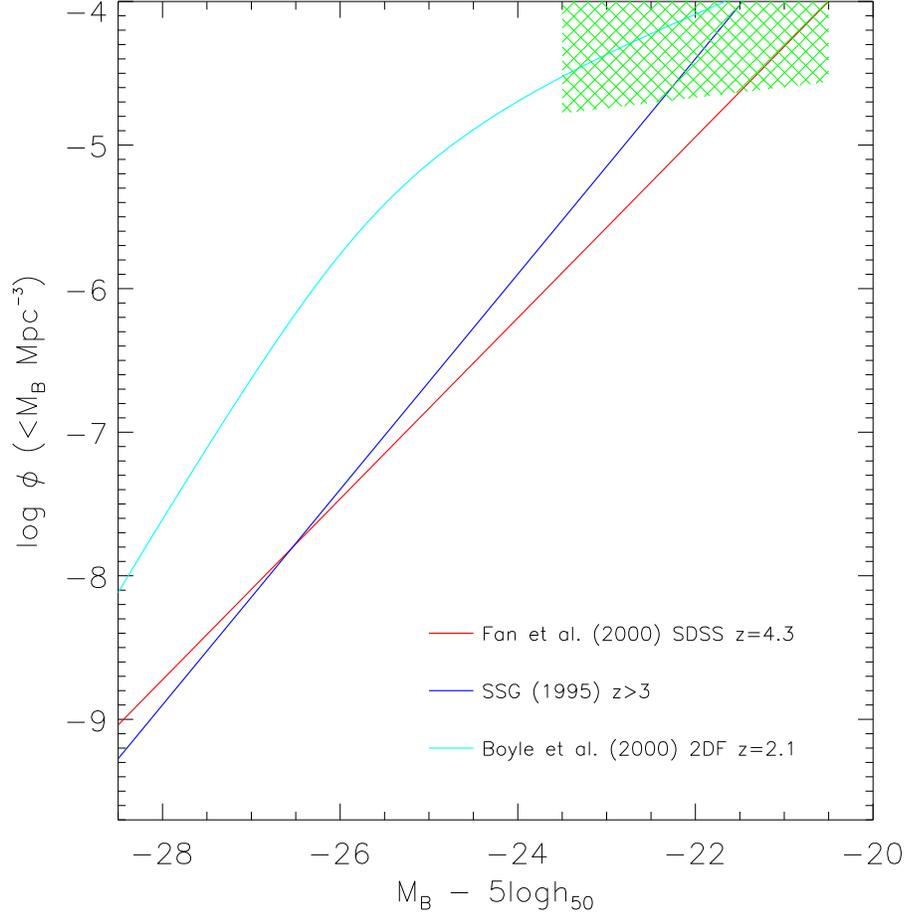}
\caption{Comparison of compilations of QSO luminosity functions with
the upper-limits we derive given that there are apparently no QSOs in
the 44 arc minute$^2$ (effectively smaller after applying the corrections
of incompleteness) region of the survey field.  The lines are the
luminosity function from \cite{Fan01} for luminous QSOs in the Sloan
Digital Sky Survey, \cite{SSG95} from the Palomar Grism Transit
Survey, and \cite{B00} z$\approx$2.1 QSOs from the 2DF survey.
The hatched region indicates approximately the region where we have
not detected any QSOs.  For comparison purposes, we have adopted the
cosmology of H$_0$=50 km s$^{-1}$ Mpc$^{-1}$ and $\Omega$=1.
\label{QSOnumlimit}}
\end{figure}

Consequently, our results strongly imply that bright galaxies did not
provide the bulk of UV photons that reionized the universe and that
their number density may have declined significantly. For reionization,
the majority of the photons must come from galaxies fainter than
our flux limit (with absolute magnitudes of $M_{AB}(1700\AA)\ga
-21$), or from brighter AGN outside the $\sim 40$ arcmin$^2$
field-of-view. Theoretically, a population of quasars with $I_{AB}\sim24$
(10-100 times fainter than the known $z>5$ Sloan Digital Sky Survey
quasars) and a surface density of around 100 per square degree could
keep the volume between $4.8<z<5.8$ ionized without appearing in our
field.  This would require a bizarre quasar luminosity function at
$z>5$, with a steep slope over the brighter parts (to provide a high
enough density of sources to cause the ionization) and then a sharp
drop, to avoid more numerous, fainter sources in fields such as ours
(Figure~\ref{QSOnumlimit}), or in the HDF.  However, for any reasonable
quasar luminosity function (i.e., one that is like those observed
for lower redshift QSOs), one would then expect several fainter AGN
at such redshifts within our field, which we do not find. Similarly,
\cite{Conti99} found no quasars in the HDF at $z>4$ to very faint
magnitudes, making this possibility extremely unlikely.

This leads to the conclusion that the bulk of the photons that reionize
the universe arise from galaxies fainter than our flux limit, those
with $M_{AB}(1700\AA)>-21$ and from galaxies with generally lower
Ly$\alpha$ equivalent widths. Our interpretation of the results of
\cite{Rhoads02} supports this, assuming that the bulk of the sources in
the \citeauthor{Rhoads02} sample are Ly$\alpha$ emitters at $z=5.7$.

\acknowledgments
We wish to thank the staff at ESO Garching and on Paranal for obtaining
these data.  More crucially, we also applaud their initiative and
efficiency in these data immediately available to us.  Specifically,
we would like to thank Martino Romaniello and Paola Sartoretti for their
dedication and hard work on our behalf. We thank the referee for their
insightful and helpful comments.

\begin{figure}
\plottwo{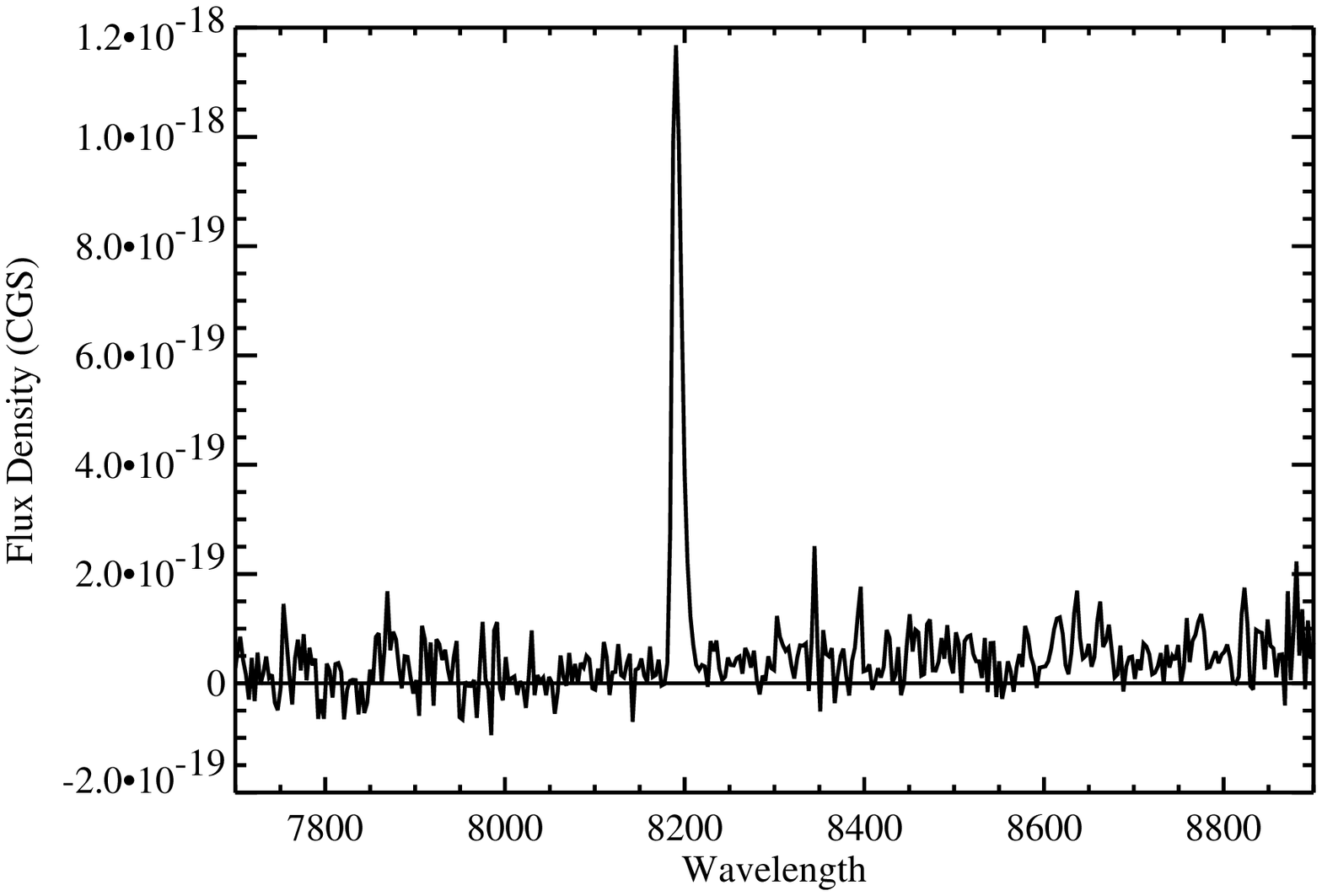}{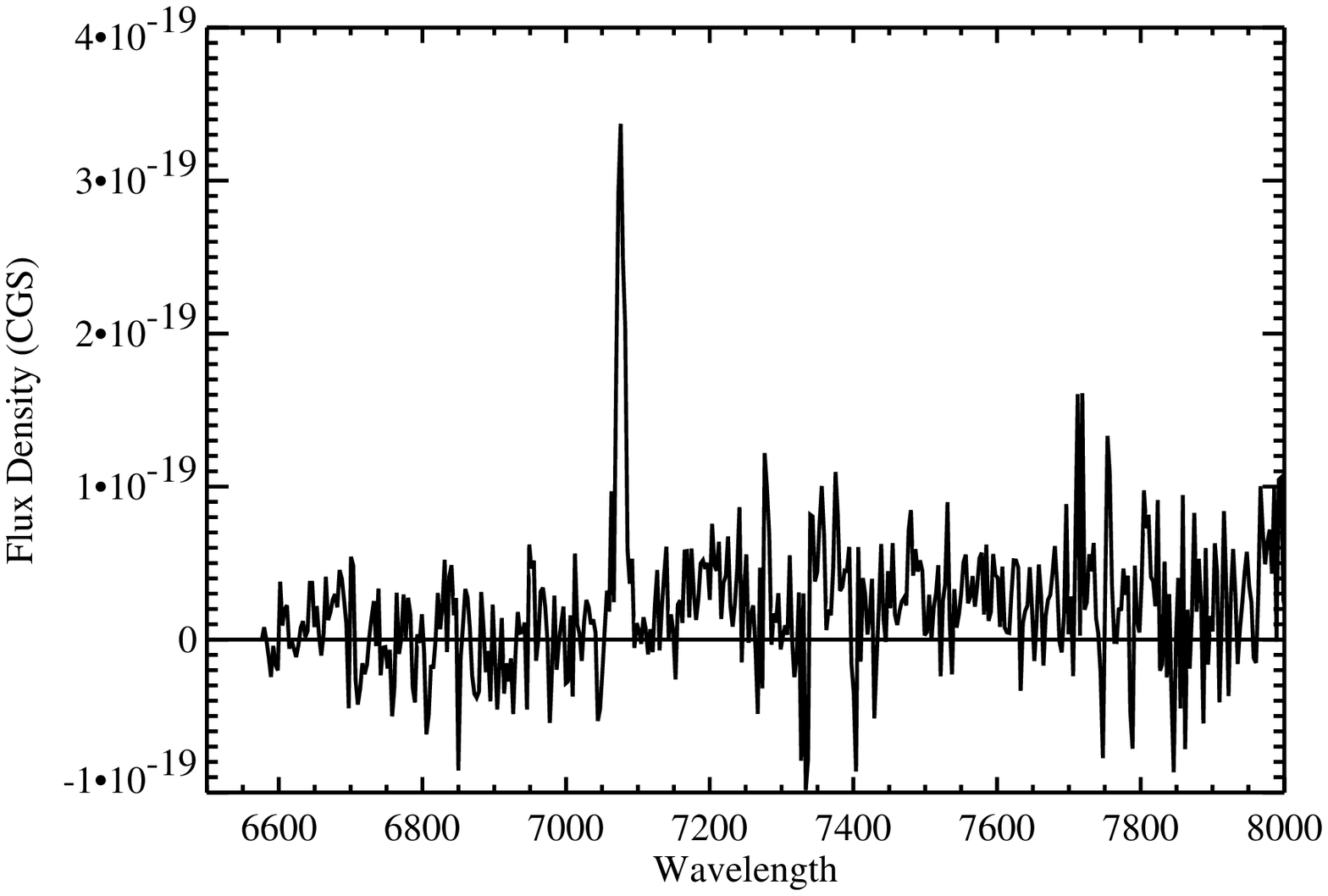}
\vskip-1.0cm
\end{figure}
\begin{figure}
\plottwo{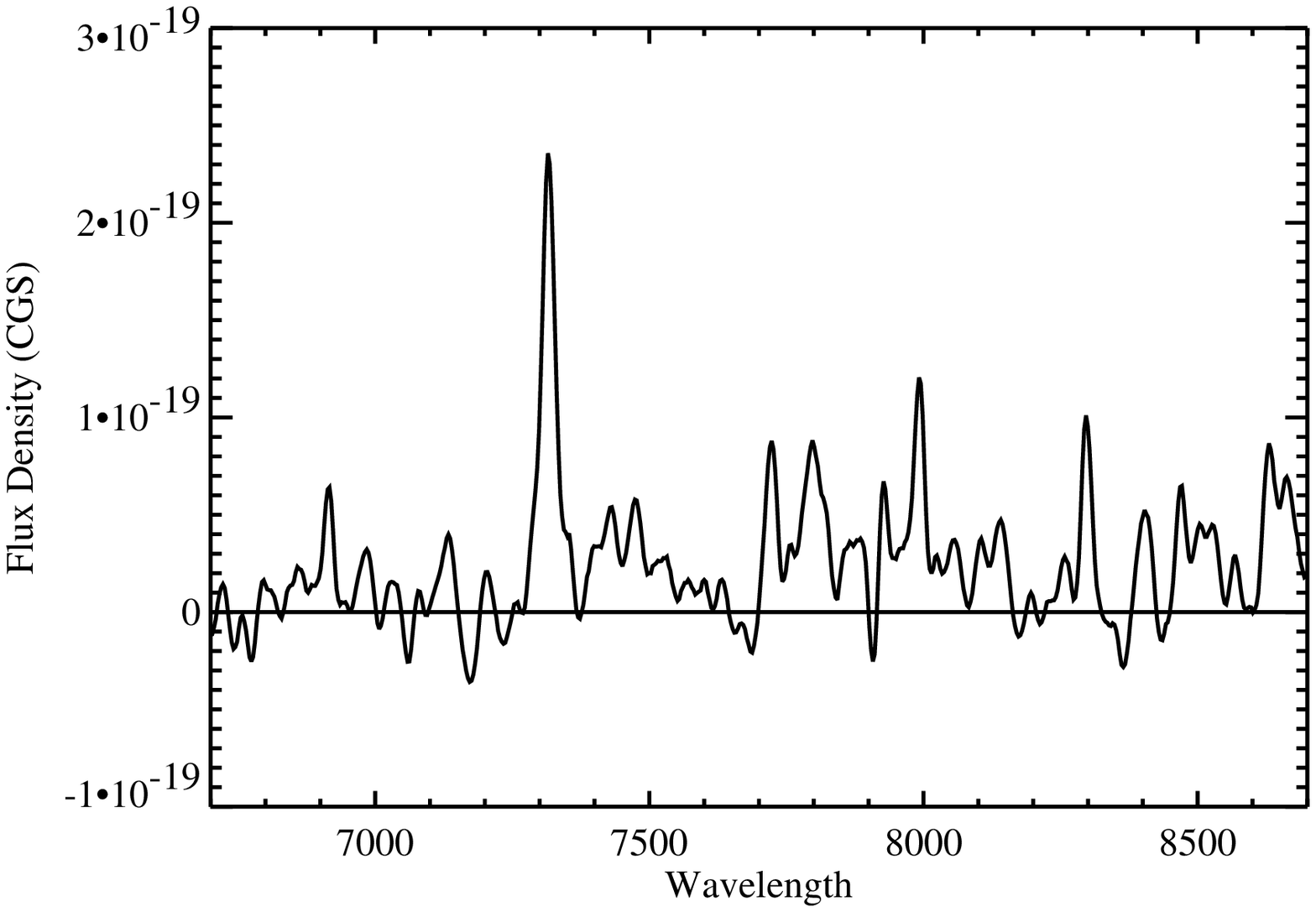}{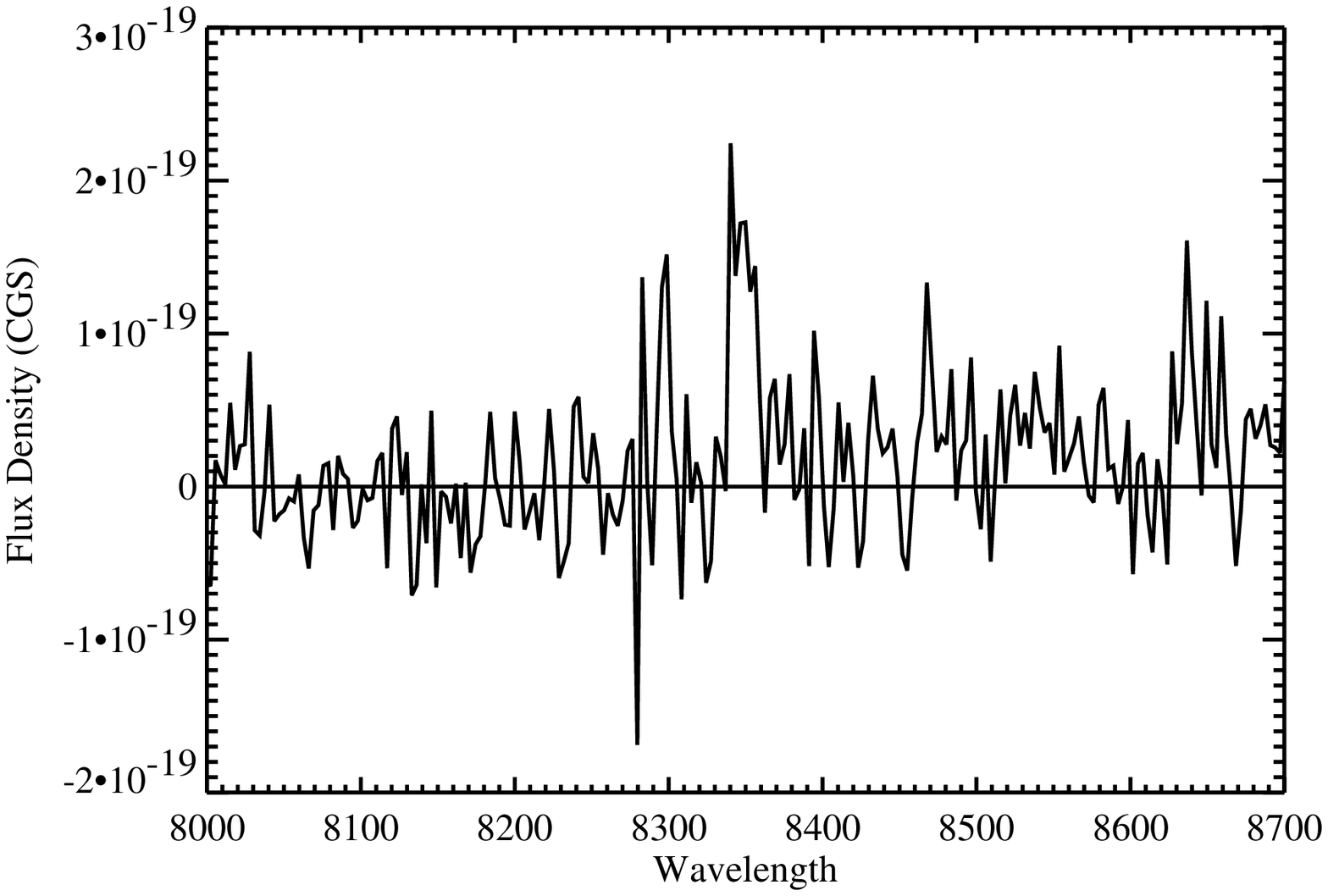}
\vskip-1.0cm
\end{figure}

\begin{thebibliography}{}

\bibitem[Becker et al. (2001)]{Becker01}Becker, R. H. et al. 2001, \aj,
122, 2850

\bibitem[Bertin \& Arnouts (1996)]{Sextractorref} Bertin, E., \& Arnouts,
S. 1996, A\&AS, 117, 393

\bibitem[Boyle et al. (2000)]{B00}Boyle, B. J., Shanks, T., Croom,
S. M., Smith, R. J., Miller, L., Loaring, N., \& Heymans, C.  2000,
MNRAS, 317, 1014

\bibitem[Calzetti et al. (2000)]{calzetti2000} Calzetti, D., Armus, L.,
Bohlin, R. C., Kinney, A. L., Koornneef, J., \& Storchi-Bergmann, T.
2000, ApJ, 533, 682

\bibitem[Charlot \& Fall (1993)]{CF93}Charlot, S., \& Fall, S. M. 1993,
ApJ, 415,580

\bibitem[Ciardi, Stoehr, \& White (2003)]{ciardi03}Ciardi, B., Stoehr,
F., \& White, S. D. M. 2003, astro-ph/0301293

\bibitem[Conti et al. (1999)]{Conti99}Conti, A., Kennefick, J. D.,
Martini, P., \& Osmer, P. S. 1999, AJ, 117, 645

\bibitem[Dawson et al. (2002)]{Dawson02}Dawson, S., Spinrad, H., Stern,
D., Dey, A., van Breugel, W., de Vries, W., \& Reuland, M. 2002, ApJ,
570, 92

\bibitem[Dey et al. (1998)]{Dey98}Dey, A., Spinrad, H., Stern, D.,
Graham, J. R., \& Chaffee, F. H. 1998, \apj,498, L93

\bibitem[Ellis et al. (2001)]{Ellis01}Ellis, R., Santos, M. R., Kneib,
J.-P., \& Kuijken, K. 2001, ApJ, 560, 119

\bibitem[Fan et al. (2001)]{Fan01}Fan et al. 2001, \aj, 122, 2833

\bibitem[Fan et al. (2002)]{Fan02}Fan et al. 2002, \aj, 123, 1247

\bibitem[Ferguson, Dickinson, \& Papovich (2002)]{FDP02}Ferguson, H. C.,
Dickinson, M., \& Papovich, C. 2002, ApJ, 569, L65

\bibitem[Fioc \& Rocca-Volmerange (1997)]{Pegase97}Fioc, M., \&
Rocca-Volmerange, B. 1997, A\&A, 326, 950

\bibitem[Hamuy et al. (1994)]{Hamuy94}Hamuy, M., Suntzeff, N. B.,
Heathcote, S. R., Walker, A. R., Gigoux, P., \& Phillips, M. M. 1994,
PASP, 106, 566

\bibitem[Hu et al. (2002)]{Hu02}Hu, E. M., Cowie, L. L., McMahon, R. G.,
Capak, P., Iwamuro, F., Kneib, J.-P., Maihara, T., \& Motohara, K. 2002,
ApJ, 568, L75

\bibitem[Iwata et al. (2003)]{iwata03} Iwata, I., Ohta, K., Tamura,
N., Ando, M., Wada, S., Watanabe, C., Akiyama, M., \& Aoki, K. 2003,
astro-ph/0301084

\bibitem[Landolt (1992)]{Landolt92} Landolt, A. U. 1992, AJ, 104, 304

\bibitem[Leitherer et al. (1999)]{Starburst99}Leitherer, C. et al. 1999
1999, ApJS, 123, 3

\bibitem[Loeb \& Barkana (2001)]{LBrev01}Loeb, A., \& Barkanna, R. 2001,
\araa, 39, 19

\bibitem[Madau, Haardt, \& Rees (1999)]{Madauradtrans99}Madau, P.,
Haardt, F. R., \& Rees, M. J. 1999, ApJ, 514, 648

\bibitem[Madau et al. (1996)]{MadauLBGs96} Madau, P., Ferguson,
H. C., Dickinson, M. E., Giavalisco, M., Steidel, C. C., \& Fruchter,
A. 1996, MNRAS, 283, 1388

\bibitem[Oke \& Gunn (1983)]{OG83} Oke, J. B., \& Gunn, J. E. 1983,
ApJ 266, 713

\bibitem[Ouchi et al. (2002)]{O02}Ouchi, M. et al. 2002, astro-ph/0202204

\bibitem[Papovich, Dickinson, \& Ferguson (2001)]{papovich01}
Papovich, C., Dickinson, M., \& Ferguson, H. C. 2001, ApJ, 559, 620

\bibitem[Rhoads \& Malhotra (2001)]{RM01}Rhoads, J. E., \& Malhotra,
S. 2001, ApJ, 563, L5

\bibitem[Rhoads et al. (2002)]{Rhoads02}Rhoads, J. E., Dey. A.,
Malhotra, S., Stern, D., Spinrad, H., Jannuzi, B. T., Dawson, S., \&
Brown, M. astro-ph/0209544

\bibitem[Schlegel, Finkbeiner, \& Davis (1998)]{Sextinction98}
Schlegel, D. J., Finkbeiner, D. P., Davis, M. 1998, ApJ, 500, 525

\bibitem[Schmidt, Schneider, \& Gunn (1995)]{SSG95}Schmidt, M., Schneider,
D. P., \& Gunn, J. E. 1995, AJ, 110, 68

\bibitem[Schneider et al. (1983)]{gunnz83} Schneider, D. P., Gunn, J. E.,
Hoessel, J. G. 1983, ApJ, 264, 337

\bibitem[Shapley et al. (2001)]{S01} Shapley, A. E., Steidel, C. C.,
Adelberger,K. L., Dickinson, M., Giavalisco, M., \& Petini, M. 2001,
ApJ, 562, 95.

\bibitem[Spinrad et al. (1998)]{Spinrad98}Spinrad, H., Stern, D., Bunker,
A., Dey, A., Lanzetta, K., Yahil, A., Pascarelle, S., Fernández-Soto,
A. 1998, AJ, 116, 2617

\bibitem[Stanway, Bunker, \& McMahon (2003)]{stanway03} Stanway, E. R., Bunker,
A. J., \& McMahon, R. G.  2003, astro-ph/0302212, MNRAS accepted

\bibitem[Steidel et al. (1999)]{SLBGs4}Steidel, C. C., Adelberger, K. L.,
Giavalisco, M., Dickinson, M., \& Pettini, M. 1999, ApJ, 519, 1

\bibitem[Stern \& Spinrad (1999)]{SS99}Stern, D. \&  Spinrad, H. 1999, PASP,
111, 1475

\bibitem[Sokasian et al. (2003)]{sokasian03}Sokasian, A., Abel, T., Hernquist,
L., \& Springel, V. 2003, astro-ph/0303098

\bibitem[Songaila \& Cowie (2002)]{SC02}Songaila, A. \& Cowie, L. L. 2002,
\aj, 123, 2183

\bibitem[Sullivan et al. (2000)]{SUV00}Sullivan, M., Treyer, M. A., Ellis,
R. S., Bridges, T. J., Milliard, B., \& Donas, J. 2000, MNRAS, 312, 442

\bibitem[Tresse et al. (1999)]{Tr99} Tresse, L., Maddox, S., Loveday,
J., \& Singleton, C. 1999, MNRAS, 310, 262

\bibitem[Weymann et al. (1998)]{Ray98}Weymann, R. J., Stern, D., Bunker,
A., Spinrad, H., Chaffee, F., Thompson, R. I., \& Storrie-Lombardi,
L. J. 1998, \apj, 505, L95

\end{thebibliography}
\end{document}